%% file: main.tex
\documentclass{article}

\usepackage[utf8]{inputenc}
\usepackage[T1]{fontenc}
\usepackage[english]{babel}
\usepackage{arxiv}
\usepackage{amsmath, amssymb}
\usepackage{microtype}
\usepackage{graphicx}
\usepackage{booktabs}
\usepackage{tabularx}
\usepackage{xcolor}
\usepackage[colorlinks=true, allcolors=blue]{hyperref}
\usepackage{url}
\usepackage{doi}
\usepackage{natbib}
\bibpunct{[}{]}{,}{n}{,}{,}
\usepackage{cleveref}
\usepackage{tikz}
\usetikzlibrary{positioning, arrows.meta, fit, backgrounds, calc, shapes.geometric}
\usepackage{pgfplots}
\pgfplotsset{compat=1.18}

\newcommand{\file}[1]{\texttt{#1}}

\input{numbers.tex}

\title{A Gated Spec-Driven Workflow: An Audit-Trail Study of Coding-Agent Work}

\author{\href{https://orcid.org/0000-0002-7232-4704}{\includegraphics[scale=0.06]{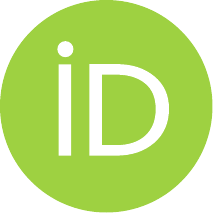}\hspace{1mm}Rodrigo Figueiredo Abdo} \\
	L-PRISMA Laboratory \\
	Federal Institute of Brasília \\
	Brasília, Brazil \\
	\texttt{rodrigo.abdo@ifb.edu.br}
	\And
	\href{https://orcid.org/0000-0002-1288-7695}{\includegraphics[scale=0.06]{orcid.pdf}\hspace{1mm}Pedro Carvalho Brom} \\
	L-PRISMA Laboratory \\
	Federal Institute of Brasília \\
	Brasília, Brazil \\
	\texttt{pedro.brom@ifb.edu.br}
}

\renewcommand{\shorttitle}{A Gated Spec-Driven Workflow}
\hypersetup{
pdftitle={A Gated Spec-Driven Workflow: An Audit-Trail Study of Coding-Agent Work},
pdfsubject={cs.SE},
pdfauthor={Rodrigo Figueiredo Abdo, Pedro Carvalho Brom},
pdfkeywords={spec-driven development, coding agents, agent scope, automated review, audit trail, case study},
}

\begin{document}
\maketitle
\begin{center}\small This work is licensed under a Creative Commons Attribution 4.0 International licence (CC BY 4.0), \url{https://creativecommons.org/licenses/by/4.0/}.\end{center}

\begin{abstract}
Spec-driven development (SDD) asks a coding agent to work from specifications, plans, and task lists rather than from the chat transcript. We study one such workflow for Claude Code, in which almost every control is an instruction carried out by the same model that it constrains. The user must approve each stage before the next begins. A reviewer subagent checks each artefact against a fixed, stage-specific defect list, and an executor subagent may only touch the files listed in its task. After each block of tasks, the orchestrator runs a file-inventory check to detect undeclared changes. We analyse the logs and transcripts of \NFeaturesExtenso{} features in \NProjetosExtenso{} projects, all run by the authors, ranging from a Fortran solver and Python programs to books and this paper. Of these, \NFeaturesConcluidasExtenso{} finished with every planned task done, including the Fortran solver feature, and \NTarefasMarcadas{} of the \NTarefas{} planned tasks were completed block by block. The executor kept to its declared files: it left only one undeclared change in the deliverable folder, and the check caught it, although its reports record other departures from the protocol. The weak point was the orchestrator, which runs and judges its own checks. It recorded undeclared changes in \NConferenciasSinalizadas{} of \NConferencias{} block checks, mostly from the authors' in-round requests and the environment, and stopped work in only \NSinalizadasParadaEfetivaExtenso{}.
\end{abstract}


\section{Introduction}

A coding agent that edits files and runs commands can complete its task and still change things that nobody asked it to change. A study by \citet{qu2026overeager} measured this behaviour on benign tasks. In a pilot of 76 paired scenarios, Claude Code running GLM-4.6 acted outside the request in 0.0\% of runs when the prompt ended with a block stating what was authorised and what was out of scope. Removing only that block raised the rate to 17.1\%. Qu et al. therefore warn that a benchmark which states the scope in the prompt measures whether the agent matches that text, not whether it can work out the boundary itself. Agents also overstate what they have completed \citep{smyth2026overclaim}, so an agent's report of what it changed is a claim rather than evidence.

Spec-driven development (SDD) has become the practitioners' answer to this problem. In SDD, developers and agents write a specification, a plan and tasks before any code is written \citep{bockeler2025sdd, github2026speckit}. Practitioners report that agents behave more reliably, but also that planning and review cost more \citep{agarwal2026sdd}. Most of the evidence on SDD comes from tool vendors \citep{agarwal2026sdd}.

This paper studies one SDD workflow whose controls are almost all instructions, carried out by the same model that they are meant to constrain. The present workflow was designed to organise complex tasks and to make the agent do what was asked and nothing else. Work passes through the stages PDR (product decision record), SPEC, PLAN, TASKS, EXECUTE, and ARCHIVE, in that order, and the user approves each artefact at a gate before the next stage starts. A reviewer subagent checks each artefact against a closed, stage-specific defect list. In EXECUTE, an orchestrator dispatches each task to an executor subagent that receives only that task and an explicit list of files it is allowed to change. After each block of tasks, the orchestrator compares an inventory of the deliverable folder with the one taken at the start of the round. Under the workflow's protocol, if this block check finds a change that the tasks executed so far did not declare, the round should stop. We call this protocol requirement the stop rule. A review episode groups the reviewer calls on one artefact until the authors send a message or a long pause occurs. Every call after the first in an episode is a re-review (\cref{sec:method}).

The controls leave written traces. The gates appear in the \texttt{Status} fields, the reviewer calls in the transcripts and logs, and the block checks in the logs, which the orchestrator writes as it decides whether to stop. We use these traces to ask where the controls held and where they did not.

\begin{description}
\item[RQ1] What do the block checks record about undeclared changes, where do those changes come from, and is the stop rule applied?
\item[RQ2] What does the reviewer find, at which stage, how is it invoked, and what does it cost?
\item[RQ3] How does the workflow behave across interruptions, across features, and when the deliverable is not software?
\end{description}

The closest prior report, by \citet{abenhaim2026specfirst}, covers one project and one refactoring carried out under a spec-first protocol, with no human review of the generated code. Compared with it, the workflow studied here adds a gate at every stage, a block check and a reviewer that works from a closed list of defects. The study also covers deliverables that are not software. We study \NFeaturesExtenso{} features in \NProjetosExtenso{} projects, which ran with all of these controls, and \NFeaturesEntregavelNaoSoftwareExtenso{} of these features delivered books, a research dossier or a research paper rather than software.

The unit of analysis is the stage mechanism rather than the completed feature, because only \NFeaturesArquivadasExtenso{} of the \NFeaturesExtenso{} features reached the ARCHIVE stage.

After reviewing related work (\cref{sec:related}), the paper describes the workflow (\cref{sec:workflow}) and the study design (\cref{sec:method}), reports the results (\cref{sec:results}), discusses their implications (\cref{sec:discussion}) and then turns to the threats to validity (\cref{sec:threats}). The main claims are listed in \cref{tab:claims} with their status, their evidence and the section that reports each.

\begin{table}[t]
\centering
\caption{The main claims of this paper, with the status and evidence of each. A descriptive claim rests on a count over the observed features, without a test. A qualitative claim rests on cases read in full in the logs and, where they exist, in the transcripts. None of the claims is comparative, because the study has no control condition.}%
\label{tab:claims}
\small
\begin{tabularx}{\textwidth}{>{\raggedright\arraybackslash}X l >{\raggedright\arraybackslash}p{4.3cm} l}
\toprule
Claim & Status & Evidence & Where \\
\midrule
Of the \NFeatures{} features, \NFeaturesEntregavelNaoSoftware{} delivered books, a research dossier or a research paper, not software & Descriptive & the deliverable of each feature (\cref{tab:features}) & \cref{sec:use} \\
The orchestrator flagged \NConferenciasSinalizadas{} of \NConferencias{} block checks as containing undeclared changes & Descriptive & orchestrator's verdict in every block-check row & \cref{sec:rq1} \\
Other undeclared changes were recorded as clean or absorbed into a new baseline & Qualitative & a reading of all block-check rows, reported as examples, not counted & \cref{sec:rq1} \\
The check stopped the round in \NSinalizadasParadaEfetivaExtenso{} of the flagged cases, once on an undeclared executor change & Qualitative & the flagged rows read in full & \cref{sec:rq1} \\
In \NRevisorProjetosComRelatorioExtenso{} projects with reviewer reports, first reviews reported a blocking finding in \NPrimeirasComBloqueante{} of \NArtefatosRevisados{} artefacts & Descriptive & \NRevisorComRelatorio{} reviewer reports in those projects & \cref{sec:rq2} \\
In P1, P2 and P5, \NRevisoesNoEpisodio{} reviewer calls were re-reviews. Of the re-reviews that came after a report without blocking findings, \NRevisoesAposSemBloqueanteComBloqueanteNovo{} reported blocking findings in the revised artefact & Descriptive & same, split into review episodes & \cref{sec:rq2} \\
Interrupted work resumed from checkboxes and logs & Qualitative & execution logs, task checkboxes and transcripts of the interrupted rounds & \cref{sec:rq3} \\
Project memory carried lessons forward in two projects & Qualitative & archiving records in the logs and the later artefacts that cite memory entries & \cref{sec:rq3} \\
Checks carried over to books, the dossier and the paper & Qualitative & comparison of installed files with the setup file, and the logs and task files of the features that delivered books, a research dossier or a research paper & \cref{sec:rq3} \\
In-round human requests, environment writes and the orchestrator deciding when rules apply strain the controls & Qualitative & the flagged rows, the reading of all rows and the reviewer calls & \cref{sec:discussion} \\
\bottomrule
\end{tabularx}
\end{table}

\section{Background and related work}\label{sec:related}

\paragraph{Spec-driven development with coding agents.} SDD treats the specification, rather than the conversation with the agent, as the artefact that governs the work. Spec Kit, Kiro and Tessl have been presented as variants of this idea \citep{bockeler2025sdd}, and open-source frameworks such as Spec Kit, OpenSpec and the BMAD Method provide templates and commands for the stages \citep{github2026speckit, openspec2026, bmad2026}. Drawing on 21 interviews and 266 repositories, a study of SDD in practice by \citet{agarwal2026sdd} found that specifications mix intent with implementation detail and that SDD is used selectively. Another study, by \citet{claudino2026epic}, scored the artefacts of 114 SDD repositories against ISO/IEC/IEEE 29148 \citep{iso29148_2018} and derived 40 practices. A comparison of six frameworks along six dimensions by \citet{macedo2026prompt} found that none of them covers all the dimensions strongly. Other accounts include a practitioner guide \citep{piskala2026sdd}, a conceptual account of SDD for agentic engineering at team scale \citep{diaz2026sdd}, a report on its use in a course \citep{tanaka2026pbl} and a case of a single engineer working with four agents \citep{vilasboas2026onedev}. The framework proposed by \citet{grabowski2026spec} combines a machine-readable specification graph, a context assembler that limits the agent's context to an ownership path, a protocol that builds vertical slices hardest first, and a drift gate. The closest prior report is a single instrumented case study by \citet{abenhaim2026specfirst}, conducted in a production codebase of 717,725 lines of TypeScript. In that study, the agent wrote a specification that went through 14 cycles of refinement against the code. The code was then implemented atomically, and it was verified against the frozen specification in 17 cycles. Over 31 audit passes, 201 defects were fixed before a human ran the program, and the specification and the raw session logs have been published.

\paragraph{Enforcement.} Controls on coding agents range from instructions that the agent is asked to follow to mechanisms that it cannot bypass. Two recent systems resemble this workflow in what they control. Spec Kit Agents adds read-only hooks for probing and hooks for validation to each stage of a Spec Kit pipeline, and was evaluated over 32 features \citep{taghavi2026speckit}. Consort uses a deterministic orchestrator, gates approved by humans, and controls that the agent runs inside but cannot bypass \citep{hartman2026consort}. Its authors distinguish three modes of enforcement. The first is enforcement by persuasion, through instructions the model may ignore. The second is enforcement by front-loaded structure, meaning strong specifications and plans followed by a trusted build. The third is enforcement through controls that the agent cannot edit. This paper takes that distinction as its lens. Apart from the reviewer's read-only tool list, which Claude Code enforces, the workflow studied here relies on enforcement by persuasion. A workflow for scientific software studied by \citet{dhruv2026scientific} makes a related point with data. There, domain experts write the specification and the plan, and every stage ends with human approval. In that study, two AI agents, an author and a reviewer, worked in a loop whose limits that workflow's orchestrator enforced in code. This loop completed about as many files as a multi-agent workflow, at about one third of the cost per file. Loops whose limits were stated only in the prompt ran past those limits, and one reached 67 iterations against a limit of 30. Other proposals enforce controls deterministically. After mining 10,008 repositories, \citet{madatha2026controlplane} found, as an indicative result, that fewer than 1\% of agent configuration files declare permission boundaries. The same work proposes hash-chained audit logs, tiered permissions enforced before the model is called, and a state machine of phases that gates feature work, with traceability from requirement to file to test. In Claim Plane, workers declare a versioned intent to change, and a deterministic control plane fails closed on any undeclared mutation that passes through it \citep{nikolaev2026claimplane}.

\paragraph{Approval and execution.} What a human approved and what an agent executed can differ. In a twelve-week study at a fintech startup, 71 of the 263 merges into main made after a non-author review gate was enforced rested solely on an approval that an agent had submitted through the tech lead's account \citep{manji2026approved}. A taxonomy by \citet{wang2026laundering} classifies six ways in which a harness executes something other than what a human approved, and measures them on the PreToolUse hook of Claude Code.

\paragraph{Scope of coding agents.} Coding agents act outside their stated scope, and benchmarks measure this by comparing files before and after a run. One benchmark compares SHA-256 snapshots \citep{qu2026overeager}, while SNARE compares the sets of files before and after a run \citep{qu2026snare}. Agents also guess where the boundaries of their actions lie when instructions are underspecified \citep{ji2026underspec}, and agents that call tools act beyond the authorisation that a user intended \citep{zhang2026authorization, zhang2026overact}. Stated constraints are often broken. The ANCHOR benchmark sets constraints on modification scope, workflow order, validation, tool use and reporting. On it, the best system reached 70.1\% atomic compliance, and only 19.0\% of tasks passed every scored item of the checklist \citep{zhu2026anchor}. Instruction files are followed literally. In 320 runs with Claude Code and Codex CLI, 90.1\% of the choices made without a corrective instruction followed a stale instruction file rather than the existing code \citep{xu2026stale}. When asked repeatedly to refactor code that was already clean, every model kept editing for several turns, and agents resumed editing after declaring the task done \citep{jiang2026overcompliant}. The workflow studied here runs the same comparison of files within the process itself, in the form of the block check that follows every block of tasks.

\paragraph{Automated review.} Prior work has tested both separate review sessions and fixed checklists. A reviewer working in a fresh session found more injected errors than a second review carried out in the same session \citep{song2026ccr}. Against a single review in the same session, or a subagent reviewer aware of the context, however, its results across runs did not differ significantly. Spec Kit offers a read-only analysis step with fixed categories and severities \citep{github2026speckit}, and LLM judges based on checklists have been proposed for text \citep{lee2025checkeval}. In inspections of requirements, scenarios aimed at specific classes of defect found more faults than generic checklists or ad hoc reading \citep{porter1995inspections}. The defect lists studied here sit between generic checklists and scenarios aimed at classes of defect: they are checklists, but each is written for one stage.

\paragraph{Position of this work.} This study differs from prior work in its data and in the controls of the workflow it examines. Studies of SDD in practice analyse the artefacts kept in repositories \citep{agarwal2026sdd, claudino2026epic}. The spec-first report by \citet{abenhaim2026specfirst} publishes its session logs, but covers only a single refactoring. We, by contrast, analyse the logs of \NProjetosExtenso{} projects, whose features run under human approval gates, a reviewer with a closed list of defects and a block check based on file inventories. Of their \NFeaturesExtenso{} features, \NFeaturesEntregavelNaoSoftwareExtenso{} delivered books, a research dossier or a research paper rather than software. Our literature search covered arXiv and the open web, but not the ACM DL or IEEE Xplore directly.

\section{The workflow}\label{sec:workflow}

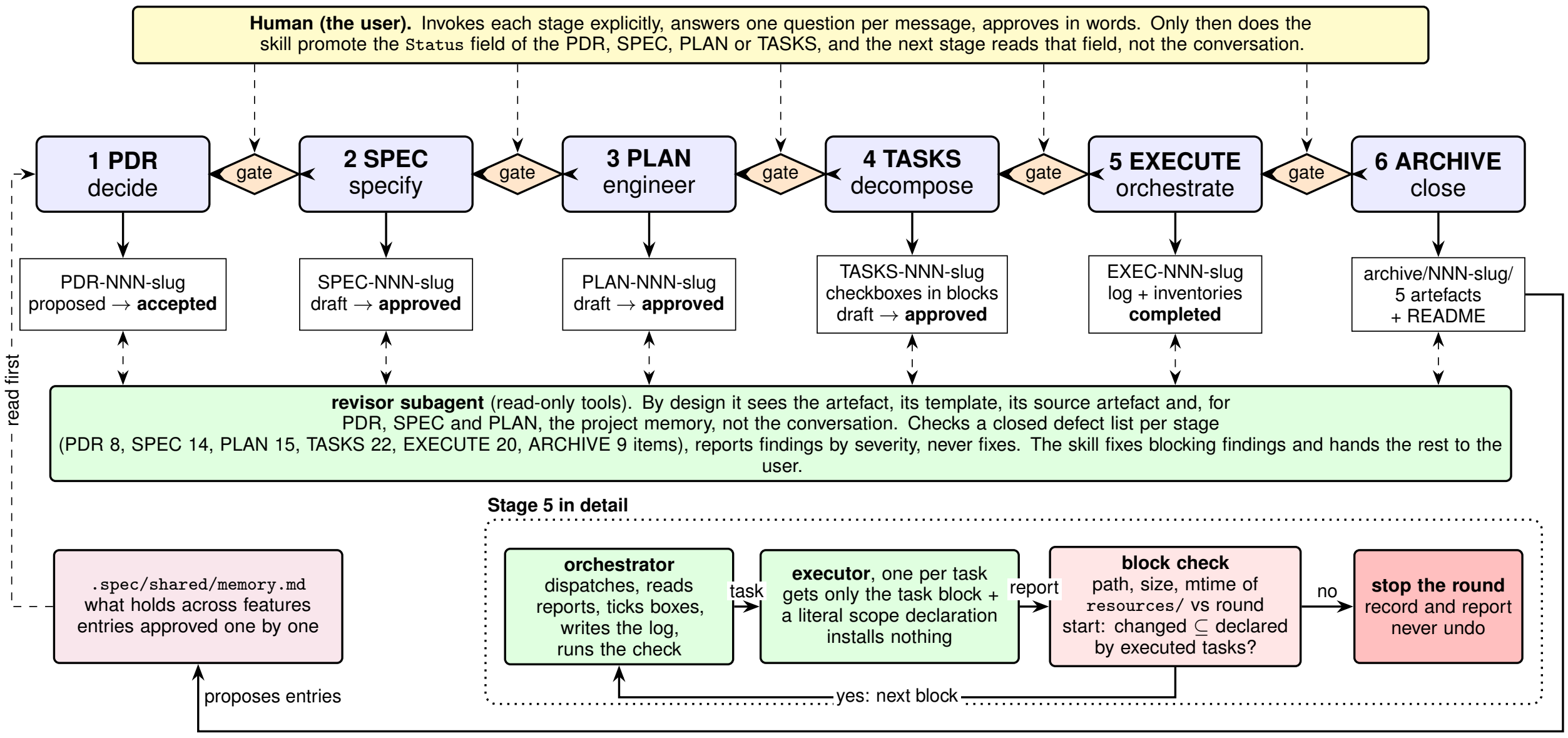
\begin{figure}[tp]
\centering
\resizebox{\textwidth}{!}{%
\begin{tikzpicture}[
  font=\sffamily\small,
  every node/.append style={execute at begin node={\hyphenpenalty=10000\exhyphenpenalty=10000}},
  stage/.style={draw, thick, rounded corners=3pt, fill=blue!8, minimum width=2.3cm, minimum height=1.0cm, align=center},
  art/.style={draw, fill=white, minimum width=2.3cm, minimum height=0.95cm, align=center, font=\sffamily\scriptsize},
  gate/.style={draw, thick, diamond, aspect=2.2, fill=orange!20, inner sep=1pt, align=center, font=\sffamily\scriptsize},
  agent/.style={draw, thick, rounded corners=2pt, fill=green!12, align=center, font=\sffamily\scriptsize},
  human/.style={draw, thick, rounded corners=2pt, fill=yellow!25, align=center, font=\sffamily\scriptsize},
  check/.style={draw, thick, rounded corners=2pt, fill=red!10, align=center, font=\sffamily\scriptsize},
  lbl/.style={font=\sffamily\scriptsize, fill=white, inner sep=1pt},
  arr/.style={-{Stealth[length=2.2mm]}, thick},
  darr/.style={-{Stealth[length=2mm]}, dashed},
]
\foreach \n/\x/\t/\v in {pdr/0/1 PDR/decide, spec/3.5/2 SPEC/specify, plan/7/3 PLAN/engineer, tasks/10.5/4 TASKS/decompose, exec/14/5 EXECUTE/orchestrate, arch/17.5/6 ARCHIVE/close}
  \node[stage] (\n) at (\x,0) {\textbf{\t}\\\v};
\node[art] (apdr) at (0,-1.6) {PDR-NNN-slug\\proposed $\to$ \textbf{accepted}};
\node[art] (aspec) at (3.5,-1.6) {SPEC-NNN-slug\\draft $\to$ \textbf{approved}};
\node[art] (aplan) at (7,-1.6) {PLAN-NNN-slug\\draft $\to$ \textbf{approved}};
\node[art] (atasks) at (10.5,-1.6) {TASKS-NNN-slug\\checkboxes in blocks\\draft $\to$ \textbf{approved}};
\node[art] (aexec) at (14,-1.6) {EXEC-NNN-slug\\log + inventories\\\textbf{completed}};
\node[art] (aarch) at (17.5,-1.6) {archive/NNN-slug/\\5 artefacts\\+ README};
\foreach \s in {pdr, spec, plan, tasks, exec, arch} \draw[arr] (\s) -- (a\s);
\foreach \a/\b in {pdr/spec, spec/plan, plan/tasks, tasks/exec, exec/arch} {
  \node[gate] (g\a) at ($(\a.east)!0.5!(\b.west)$) {gate};
  \draw[arr] (\a) -- (g\a); \draw[arr] (g\a) -- (\b);
}
\node[human, text width=17cm] (human) at (8.75,1.85) {\textbf{Human (the user).} Invokes each stage explicitly, answers one question per message, approves in words. Only then does the skill promote the \texttt{Status} field of the PDR, SPEC, PLAN or TASKS, and the next stage reads that field, not the conversation.};
\foreach \a in {pdr, spec, plan, tasks, exec} \draw[darr] (human.south -| g\a) -- (g\a.north);
\node[agent, text width=19.2cm, minimum height=1.0cm] (rev) at (8.75,-3.45) {\textbf{revisor subagent} (read-only tools). By design it sees the artefact, its template, its source artefact and, for PDR, SPEC and PLAN, the project memory, not the conversation. Checks a closed defect list per stage\\(PDR 8, SPEC 14, PLAN 15, TASKS 22, EXECUTE 20, ARCHIVE 9 items), reports findings by severity, never fixes. The skill fixes blocking findings and hands the rest to the user.};
\foreach \a in {apdr, aspec, aplan, atasks, aexec, aarch} \draw[darr, {Stealth[length=2mm]}-{Stealth[length=2mm]}] (\a.south) -- (\a.south |- rev.north);
\node[agent, text width=2.8cm, minimum height=1.5cm] (orch) at (6.6,-5.75) {\textbf{orchestrator}\\dispatches, reads reports, ticks boxes, writes the log, runs the check};
\node[agent, text width=3.2cm, minimum height=1.5cm] (execu) at (10.2,-5.75) {\textbf{executor}, one per task\\gets only the task block + a literal scope declaration\\installs nothing};
\node[check, text width=3.1cm, minimum height=1.5cm] (inv) at (14,-5.75) {\textbf{block check}\\path, size, mtime of \texttt{resources/} vs round start: changed $\subseteq$ declared by executed tasks?};
\node[check, text width=2.0cm, minimum height=1.5cm, fill=red!25] (stop) at (17.5,-5.75) {\textbf{stop the round}\\record and report\\never undo};
\draw[arr] (orch) -- node[lbl, above=2pt]{task} (execu);
\draw[arr] (execu) -- node[lbl, above=2pt]{report} (inv);
\draw[arr] (inv) -- node[lbl, above=2pt]{no} (stop);
\draw[arr] (inv.south) -- ++(0,-0.4) -| node[lbl, pos=0.25]{yes: next block} (orch.south);
\begin{scope}[on background layer]
\node[draw, dotted, thick, rounded corners, fit=(orch)(execu)(inv)(stop), inner sep=7pt, yshift=-2pt, minimum height=2.45cm, label={[font=\sffamily\scriptsize\bfseries, anchor=south west, inner sep=1pt]north west:Stage 5 in detail}] (s5) {};
\end{scope}
\node[human, fill=purple!10, text width=3.6cm, minimum height=1.5cm] (mem) at (1.0,-5.75) {\texttt{.spec/shared/memory.md}\\what holds across features\\entries approved one by one};
\draw[darr] (mem.west) -- ++(-0.55,0) |- node[lbl, pos=0.25, rotate=90]{read first} (pdr.west);
\draw[arr] (aarch.east) -- ++(0.5,0) |- ($(s5.south)+(0,-0.35)$) -| node[lbl, pos=0.75, right=1pt]{proposes entries} (mem.south);
\end{tikzpicture}}
\caption{The pipeline as designed. Each stage starts only when the user invokes it, and only if the previous artefact carries the required \texttt{Status}. The reviewer audits each artefact against the closed defect list for its stage. In stage 5, EXECUTE, the orchestrator dispatches each task to its own isolated executor. After each block, it compares an inventory of the deliverable folder with the inventory taken at the start of the round. Any change outside the files declared by the tasks already executed must stop the round. Almost every control is carried out by an agent that follows instructions.}%
\label{fig:pipeline}
\end{figure}

The workflow is installed from a single Markdown file: the user gives the agent \file{SETUP-SDD.md}, which creates the folders, six skills, five templates, the project memory and two subagents \citep{abdo2026template}. The complete pipeline, as designed, is shown in \cref{fig:pipeline}.

\paragraph{Stages and gates.} Each of the six stages is a skill that runs only when the user explicitly invokes it. Its artefact must then pass a gate before the next stage can start. The PDR, SPEC, PLAN and TASKS each carry a \texttt{Status} field, which the skill promotes only after the user has approved in words. Once every task has been ticked, the orchestrator marks the execution log as completed. The PDR stage interviews the user, asking one question per message, and records one decision, together with its discarded alternatives, its consequences and an observable review trigger. The PDR itself is an adaptation of the architecture decision record \citep{nygard2011adr, buchgeher2023adr}. Before proposing anything, the SPEC stage does its research, and it keeps a feasibility table in which every technical claim is confirmed in a source that has been opened. The PLAN stage reads the deliverable folder and writes a file map, a list of side effects, and steps that each carry their own verification, and it groups independent steps into waves. The TASKS stage then writes atomic tasks for an executor that has seen nothing else, and it groups them into blocks whose tasks share no dependency, no file they write and no resource. A \texttt{Parallelism} field starts as \emph{no} and changes only if the user says so in words. Only the EXECUTE stage writes to \file{resources/}, the root of the deliverable. Last, the ARCHIVE stage moves the artefacts away and proposes entries for the project memory, which it is the only stage allowed to write.

\paragraph{Reviewer.} Once a skill has written its artefact, it invokes the \texttt{revisor} subagent, which we call the reviewer, and passes it the artefact's path and stage. The reviewer's tools are read-only. It reads the artefact, its template and its source artefact, and, for the PDR, SPEC and PLAN, the project memory as well. The reviewer checks the artefact against the closed list of defects for that stage, but it does not fix the defects it finds. Each finding names the item it refers to and is given one of three severities: blocking, serious or observation. The skill fixes the blocking findings before delivering the artefact, hands the remaining findings to the user, and does not ask for a second review. The reviewer's instructions state that ``no defect found'' is a legitimate answer and is the answer expected most of the time.

\paragraph{Executor and scope declaration.} In the EXECUTE stage, the main agent, which we call the orchestrator, dispatches each task to a new \texttt{executor} subagent. The executor receives the task block, a fixed scope paragraph and nothing else. That same paragraph, copied unchanged, ends every dispatch and limits the executor to the files listed in the task. The executor installs nothing and stops whenever it is in doubt. When it sees something wrong, it records the problem but does not fix it. We call this record of things seen but left unfixed the valve. The execution log also contains deviation rows, in which the orchestrator records anything that departed from the plan. The workflow has no rule for amending the task file during a round, yet orchestrators added numbered amendments, which the authors approved.

\paragraph{Block check and stop rule.} The block check looks for changes in \file{resources/} that no executed task declared. Before each execution round, the orchestrator records the path, size, and modification time of every file in \file{resources/}. After every block, it takes a new inventory and compares it with the baseline from the start of the round. It then checks whether every changed, created, or removed file belongs to the union of the files declared by the tasks executed so far in the round. Under the workflow's protocol, any change outside that union should be recorded and should stop the round (the stop rule). The orchestrator is also instructed not to ``undo'' the change on its own.

The workflow declares an explicit limit of this check: it does not see writes outside \file{resources/}. Two further limits follow from the design. First, the comparison is cumulative, so a later block that changes a file already declared by an earlier task in the same round can go unnoticed. Second, the orchestrator both runs the check and records its outcome, so the check is only as reliable as the orchestrator's adherence to the protocol.

\paragraph{Design validation.} Before the use reported in this paper, the authors tested the design on a clean installation with three end-to-end features of increasing difficulty \citep{abdo2026template}. The commit that records these features (ac7f636) lists the defects they revealed and the rules that were added in response. These rules include the default value of the \texttt{Parallelism} field and the declared limit of the block check. The setup file has not changed since 10 September 2026.

\section{Study design}\label{sec:method}

Following the case-study guidelines of \citet{runeson2009case}, we conducted an observational multiple-case study. The work was not changed for the study, and we analysed only work artefacts that already existed, which these guidelines call third-degree data.

\paragraph{Case selection.} The cases are all the features run with the final version of the workflow between 11 September and the end of October 2026. They comprise \NFeaturesExtenso{} features in \NProjetosExtenso{} projects and form a census rather than a sample. P6 ran on another of the authors' machines, and the authors supplied its artefacts. The other projects ran on the authors' main machine. In all \NProjetosExtenso{} installations, the skills, subagents and templates match the blocks of the published setup file. The projects are all identified only by their scope (\cref{tab:features}).

\paragraph{Units of analysis.} Each unit is one stage mechanism: a gate passage, a reviewer call grouped into review episodes (defined below), a block check or a task record. Only \NFeaturesArquivadasExtenso{} of the \NFeaturesExtenso{} features reached the ARCHIVE stage, so conclusions about whole features are beyond the scope of this study.

\paragraph{Data sources and coverage.} The coverage of each source differs from project to project. The first source is the set of artefacts of each feature, including the execution logs and the saved inventories. The second is the Claude Code transcripts, which record every reviewer call with its prompt, report, token count and duration. Main transcripts, those of the main agent's sessions, survive for P1, P2 and P5, although for P1 they cover only its first feature. For P4 and for the later sessions of P1, only the metadata of their subagents survive, and P3 and P6 have no transcript data at all. P6 ran on another machine, and its transcripts were not available to us. P7 has no transcript either. Its reviewer calls, findings, token counts and run times come from an event record that the orchestrator kept at each gate and each reviewer call. Because the review episodes and \cref{fig:calls} depend on the authors' messages in the transcripts, P7 enters neither. The resulting coverage of the reviewer data is shown in \cref{tab:coverage}. Where transcripts exist, 21 of the 4,483 model responses ran on Claude Opus~5.5, all in the main sessions of P2 and P5, and all other responses of the orchestrator and the subagents ran on Claude Opus~5 (\texttt{claude-opus-5}). The third source is the history of the workflow repository and, in P2 and P5, the git history of the deliverable, which holds the authors' own commits.

\begin{table}[t]
\centering
\caption{Coverage of the reviewer data, by project. Known: reviewer subagents recorded in the transcript metadata, plus reviews sent to a resumed reviewer. For P7, Known counts the reviewer calls in the event record that the orchestrator kept during the feature. With report: reports recovered from the transcripts or, for P7, from that event record. Without report: known calls for which no report was recovered. P3, P6 and P7 have no transcript data, and for P4 only the metadata survive.}%
\label{tab:coverage}
\small
\begin{tabular}{l r r r}
\toprule
Project & Known & With report & Without report \\
\midrule
\input{tables/coverage.tex} 
\bottomrule
\end{tabular}
\end{table}

\paragraph{Extraction and chain of evidence.} Python scripts read the sources and write CSV files. Further scripts then generate from these files the data tables, \cref{fig:calls} and every count across features and projects, except the counts obtained by reading the logs, transcripts and artefacts. Numbers within a single case, such as the time between checks, come from reading the logs and transcripts. On every reviewer report, we checked the script that counts the report's findings: the script's counts per severity had to add up to the number of items listed under the severity headings. On two reports we also counted the findings by hand. We read the flagged block checks in full in the logs, and also read the P2 case in the transcript and the git history. Our coding of these block checks is recorded in the extraction script and its input files, and \cref{tab:flagged} is generated from the resulting data.

\paragraph{Review episodes.} In P1, P2 and P5, the projects with transcripts, a review episode starts at the first review of an artefact. A later review of that artefact starts a new episode if the authors have written any message since the previous review, or if more than three hours have passed. We recomputed the episodes with gaps of one and six hours, and the counts reported below did not change. Messages from the authors are detected from the transcripts. Tool notifications, slash commands, messages from other agent sessions, interruption markers and automatic context summaries are excluded.

\paragraph{Researcher bias.} The authors are the designers of the workflow, its only users and the subjects of the cases. Claude Code carried out the analysis, including the coding of origins and episodes, under the authors' direction. It worked on logs that Claude Code itself had written, and there was no second human coder. To limit this bias, scripts compute the counts across features and projects from the raw files, except those obtained by reading the logs, transcripts and artefacts. The manuscript also went through an evidence audit, a technical review and later review rounds, all carried out by separate agent sessions without access to the analysis. These reviews found claims that the data did not support, including our first readings of the repeated reviewer calls and of the P2 figure case (\cref{sec:rq2,sec:rq1}). The analysis reported here is the corrected one. The agents that reviewed the manuscript are models too, which limits what they can catch.

\section{Results}\label{sec:results}

\subsection{Observed use}\label{sec:use}

\begin{table}[t]
\centering
\caption{The features studied. Tasks: tasks planned and tasks checked as done. Blocks: blocks of tasks in the task file. Rounds: execution rounds, one for each start or resumption. Checks: block-check rows in the log. Dev.: deviation rows. Valve: rows in the record of things seen and not fixed. Log status: the status of the feature as recorded in its execution log.}%
\label{tab:features}
\footnotesize
\setlength{\tabcolsep}{4pt}
\begin{tabular}{l l r r r r r r r l}
\toprule
Feature & Deliverable & \multicolumn{2}{c}{Tasks} & Blocks & Rounds & Checks & Dev. & Valve & Log status \\
\cmidrule(lr){3-4}
 & & plan. & done & & & & & & \\
\midrule
\input{tables/features.tex} 
\bottomrule
\end{tabular}
\end{table}

Across the \NFeaturesExtenso{} features, the task files planned \NTarefas{} tasks in \NBlocos{} blocks, and \NTarefasMarcadas{} of these were checked as done over \NRodadas{} execution rounds (\cref{tab:features}).

Of these features, \NFeaturesConcluidasExtenso{} were completed and \NFeaturesArquivadasExtenso{} were archived. Of the other four, two were interrupted and two are still in execution (\cref{tab:features}), but the records they have already produced are sufficient for the analysis and the preliminary conclusions of this study. In \NFeaturesEntregavelNaoSoftwareExtenso{} features, the deliverable was a book, a research dossier or a research paper.

The \texttt{Parallelism} field remained at \emph{no} in every task file except the one for P7, where the authors asked in words for parallel execution wherever it was pertinent. In P1, P2 and P5, the projects with main transcripts, the authors chose serial execution, also in words.

The valve was used often, but one case shows its limit. According to the P2 log, the executor noted in four tasks of the same round that two figures in a chapter were not referenced in the text. Each time, the orchestrator copied the entry into the log's findings, yet nobody acted on the problem until the authors found it while reading the chapter.

The P2 log also records commits and online edits that the authors made to the deliverable while rounds were running, one of them while a task was writing to the same chapter. This happened even though the feature's own specification had stated that the deliverable would change only through tasks.

\subsection{RQ1: block checks and the stop rule}\label{sec:rq1}

The logs contain \NConferencias{} block-check rows, and every executed block appears in at least one of them. Coverage is uneven, however. In P4, groups of blocks were checked together, some amended tasks had no check of their own, and the log notes that its check times are approximate. Only round-level inventories were saved, \NInventariosSalvos{} files in all, so the block checks cannot be recomputed. They are the orchestrator's own record.

We use two measures. The first is mechanical: a row counts as flagged when the orchestrator's own verdict on whether all changes were declared does not begin with ``yes''. By this measure, the orchestrator flagged \NConferenciasSinalizadas{} rows, which we read in full and coded (\cref{tab:flagged}). The second measure is a reading of all the rows. It finds further undeclared changes in rows recorded as ``yes''. We describe these with examples below but do not count them, because whether they count depends on judgements, such as whether a change the authors authorised counts as undeclared, that a single coder cannot settle.

\begin{table}[t]
\centering
\caption{The block checks that the orchestrator itself flagged, with the origin of each undeclared change and what happened afterwards. We coded these cases from the logs and, for P2, also from the transcript and the git history.}%
\label{tab:flagged}
\small
\begin{tabularx}{\textwidth}{l >{\raggedright\arraybackslash}X >{\raggedright\arraybackslash}X}
\toprule
Feature & Origin of the change & What happened next \\
\midrule
\input{tables/flagged.tex} 
\bottomrule
\end{tabularx}
\end{table}

Of the flagged rows, \NSinalizadasSincronizacaoExtenso{} came from the cloud-sync client. In P6 it wrote temporary files, and in P7 it set the modification times of scripts back and reverted a build product. In \NSinalizadasPedidoExtenso{} cases, the orchestrator changed files outside any task because the authors had requested or authorised it mid-round. One of these also included a date change that came from the authors' own commits. For one case in this group the logs disagree: the deviation row cites no request, whereas the block-check row and the contamination table do. In P3, a verification command rebuilt files that its task did not list, even though the workflow requires each task to declare what its verification writes. Finally, in P6-F2, the executor once ran a syntax check on its own initiative and left byte-code caches behind.

The check stopped the round in \NSinalizadasParadaEfetivaExtenso{} cases, half in P6 and half in P7. In P6, one stop was caused by a sync-client file. The other was caused by the executor's syntax check, and the round continued only after the authors approved deleting the caches. In P7, both stops were caused by sync-client changes. Each time, the work resumed in a new round, under the authors' order to go on without asking at each step. In two further P6 cases, the round stopped for a different reason, a failed task verification, and the sync-client files were recorded as external. In P2, the orchestrator logged a stop and, seconds later, ran a script to rewrite that verdict as a pass. The script edited a different row by mistake, so the stop survived in the log. The next block was then checked four minutes later in the same round, with no message from the authors in between. In the two P1-F2 cases and the P3 case, the orchestrator recorded a reason and carried on. In the last P6 case, it deleted the sync-client files without asking, and the round then paused for the authors.

A reading of every row reveals undeclared changes that the orchestrator recorded as ``yes''. In P6, after the authors had approved deleting one batch of sync-client files, the orchestrator deleted such files without asking in two rounds, against the rule that it must not undo changes on its own. One of those checks is among the flagged rows above, and the other was recorded as ``yes''. Also in P6, the orchestrator regenerated pipeline stages outside any task and moved files left behind by a test battery. Other ``yes'' rows list a folder that a verification created and that its task listed among its resources but not among its files. They also list files written when the orchestrator, under an amendment, used a headless browser to carry out the steps a user would normally perform. Further rows list changes that the orchestrator made with the authors' authorisation in P4 and P6, and a chapter in P5 whose date changed through the authors' own git pull. In P2, the authors made an online edit to a chapter that a running task had declared, and this edit passed the cumulative check unseen, as the design allows.

The logs also show at least three occasions on which the orchestrator took a new inventory baseline that absorbed changes made outside tasks. In P1-F2, it created a file at the authors' request and amended the baseline to include it. In P5, it moved files out on the authors' order and took a new baseline. In P6, the sync client sent a whole folder of the deliverable to the trash, and the orchestrator restored the files and took a new baseline.

Within \file{resources/}, the executor generally limited its writes to the files declared by its current tasks. We observed exceptions: the syntax check above, an extra inspection in P6-F1 described below, and an extra compilation in P7 that created and deleted a cache folder. The executor also departed from other protocol constraints. In P4, its task instructed it to run an installation, which the workflow reserves for the orchestrator. The log records this as a defect of the task file. In P6-F1, it ran an extra inspection that recreated and deleted a folder. These actions touched \file{resources/} but did not necessarily leave durable inventory differences: the installation wrote to a declared path, and the inspection and compilation left no net change.

In P6-F2, the executor deleted temporary folders without the confirmation that deletions require. It also terminated two of its own processes, read another task's block, and refactored code that nobody had asked it to change. Across several features, it wrote scratch files, caches, and downloads outside \file{resources/}, which only its reports reveal. Because we did not code all \NDesvios{} deviation rows, these are illustrative examples rather than a count.

The P2 figure case shows an approval gate being overridden despite a warning. The orchestrator asked the authors whether to keep the original figures of a chapter or to replace them with openly licensed ones. The authors replied by asking to redo ``the changes related to the author, etc.'' The orchestrator took this reply as consent and added four tasks to the approved task file during execution. When asked to audit them, the reviewer reported as a serious finding that no one had approved the new tasks and that one of them left a decision to the executor. The orchestrator edited the tasks and dispatched them within two minutes, without a new review and without asking the authors. After the replacement, the authors objected and ordered the reversal that the check later flagged. That execution round also closed without the review of the execution log that the skill requires.

These results describe what the checks recorded, not what the controls prevented. Since there is no condition in which the scope declaration or the check is absent, the scarcity of out-of-scope edits by the executor cannot be credited to either of them.

\subsection{RQ2: what the reviewer finds and how it is invoked}\label{sec:rq2}

\begin{table}[t]
\centering
\caption{Reviewer reports by stage, covering the \NRevisorProjetosComRelatorioExtenso{} projects that have reviewer reports. EXEC\@: the EXECUTE stage. Artefacts: distinct artefacts reviewed. Calls: reviewer calls that returned a report. Clean: reports stating ``no defect found''. Blocking, Serious, Observations: findings of each severity. First: artefacts whose first review had at least one blocking finding. Tokens: subagent tokens in millions. This is a lower bound, because \NRevisorSemTokens{} calls have no token count.}%
\label{tab:reviewer}
\small
\begin{tabular}{l r r r r r r r r}
\toprule
Stage & Artefacts & Calls & Clean & Blocking & Serious & Observations & First & Tokens (M) \\
\midrule
\input{tables/reviewer.tex} 
\bottomrule
\end{tabular}
\end{table}

The \NRevisorComRelatorio{} reports cover \NArtefatosRevisados{} artefacts and contain \NRevisorAchados{} findings, of which \NRevisorBloq{} are blocking (\cref{tab:reviewer}). The reviewer reported a blocking finding in the first review of \NPrimeirasComBloqueante{} of the \NArtefatosRevisados{} artefacts, including the first review of every specification and every task file. The blocking findings are concentrated on a few items of the defect lists (\cref{tab:blocking}). The most frequent are technical claims in a specification that have no line in the feasibility table, subjective acceptance criteria, and task verifications that cannot be run as written. Also among them are PDRs that hold two decisions, specifications that leave a decision of the PDR uncovered, and tasks that leave a decision to the executor. Without ground truth, we cannot say how many findings were real defects or how many defects the reviewer missed.

\begin{table}[t]
\centering
\caption{Blocking findings by stage and by the item of the defect list that they cite. Item: the label of the item in the list for its stage. Codes: the number of citations of the item. Of the \NRevisorBloq{} blocking findings, \NBloqueantesMulticodigo{} cite more than one item, so the Codes column sums to \NBloqueantesCodigos{}. Item U2, the check against the project memory, applies to the PDR, SPEC and PLAN stages. EXEC\@: the EXECUTE stage.}%
\label{tab:blocking}
\small
\begin{tabular}{l l l r}
\toprule
Stage & Item & Defect & Codes \\
\midrule
\input{tables/blocking.tex} 
\bottomrule
\end{tabular}
\end{table}

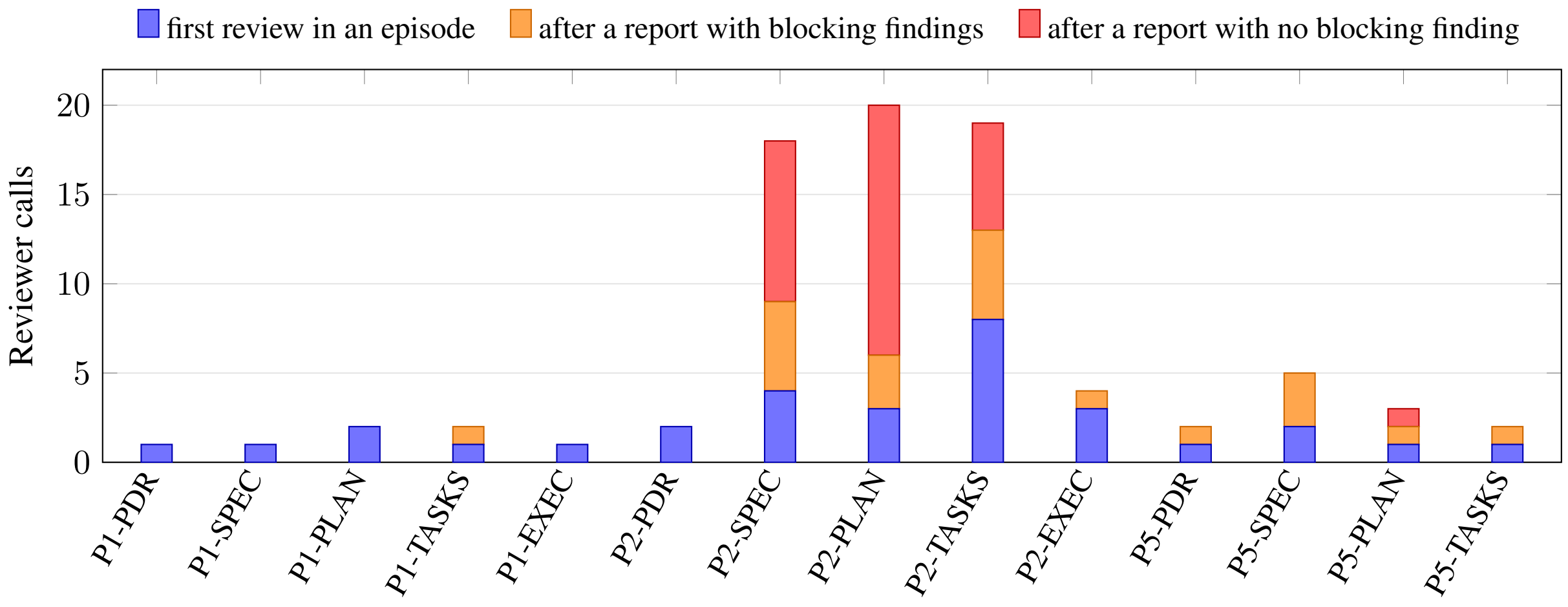
\begin{figure}[t]
\centering
\begin{tikzpicture}
\begin{axis}[
  ybar stacked, bar width=9pt,
  width=\textwidth, height=5.6cm,
  ylabel={Reviewer calls},
  xtick=data, xticklabels from table={tables/calls_by_trigger.dat}{artifact},
  x tick label style={rotate=60, anchor=east, font=\footnotesize},
  ymin=0, enlarge x limits=0.04,
  legend style={at={(0.5,1.03)}, anchor=south, font=\footnotesize, draw=none, legend columns=3, /tikz/every even column/.append style={column sep=8pt}},
  legend cell align=left,
  grid=major, grid style={gray!20}, xmajorgrids=false,
]
\addplot+[fill=blue!55, draw=blue!70!black] table[x expr=\coordindex, y=first] {tables/calls_by_trigger.dat};
\addlegendentry{first review in an episode}
\addplot+[fill=orange!70, draw=orange!80!black] table[x expr=\coordindex, y=after_blocking] {tables/calls_by_trigger.dat};
\addlegendentry{after a report with blocking findings}
\addplot+[fill=red!60, draw=red!70!black] table[x expr=\coordindex, y=after_zero] {tables/calls_by_trigger.dat};
\addlegendentry{after a report with no blocking finding}
\end{axis}
\end{tikzpicture}
\caption{Reviewer calls per artefact in the projects with transcripts, grouped by what preceded each call. The skill asks for one review per artefact, with blocking findings fixed and the rest handed over to the user. Every call after the first in an episode goes beyond that instruction, and most such calls are in P2.}%
\label{fig:calls}
\end{figure}

In \NRevisorComRelatorio{} calls, the reviewer returned a clean report \NRevisorLimpasExtenso{} times, even though its own instructions describe a clean report as the expected answer most of the time. In P1, P2 and P5, the calls form \NEpisodios{} review episodes, and \NEpisodiosIniciadosPorAutor{} of these began after a message from the authors. Within the episodes, \NRevisoesNoEpisodio{} calls were re-reviews, which go beyond what the skill asks for (\cref{fig:calls}). Of these re-reviews, \NRevisoesAposRelatorioSemBloqueante{} came after a report that already contained no blocking finding. They used \NFracaoTokensAposSemBloqueante\% of the reviewer's tokens in those projects, and \NRevisoesAposRelatorioSemBloqueanteP{} of them took place in P2. In \NRevisoesAposSemBloqueanteComBloqueanteNovo{} of the re-reviews that came after a report without blocking findings, the reviewer still reported blocking findings in the revised artefact. Because the artefact changed between calls, we cannot tell how many of these findings the earlier review had missed.

In the P2 transcript, the longest run of re-reviews covered three artefacts, each reviewed six or seven times within two hours. This run followed an instruction from the authors to ``carry on with the adjustments and checks''. The orchestrator announced that it would revise the specification, the plan and the tasks, calling the reviewer after each one. It then went on reviewing even after reports without blocking findings. The same pattern had appeared three days earlier, when a plain approval was followed by eight reviews of the specification. Across P2, \NPSemMsgAutor{} of the \NPChamadas{} reviewer calls were made with no message from the authors since the previous call.

The reviewer was not always kept independent from the orchestrator's running narrative. Of the \NPromptsAgente{} reviewer prompts recorded in the transcripts, \NPromptsComContexto{} included text written by the orchestrator in addition to the artefact path, such as summaries of earlier findings and claims that they had been fixed. One re-review was sent to a resumed reviewer, which retained earlier context. Independence from the conversation is a primary motivation for a separate reviewer \citep{song2026ccr}. In the transcripts, that independence held only in the remaining calls, a little over half of those recorded.

\subsection{RQ3: interruptions, memory and deliverables that are not software}\label{sec:rq3}

The workflow kept long and interrupted work recoverable. According to the logs, P2 ran in eight execution rounds over fifteen days and P6-F1 in nine rounds over six days, with each log adding a row for every resumption. The surviving transcripts do not cover the last four rounds of P2. In each of P1, P2 and P5, one session ran out of context and continued from an automatic summary, and the work then went on from the task checkboxes, the execution log and the \texttt{Status} fields. In P5, by contrast, a round was left open. Its log still reads \emph{in execution} although no round is running, which is the dead-round state that the workflow tells the next run to close with the user. A later session asked where the work had stopped, but it did not close the round.

The project memory carried lessons forward in two projects. According to the archive README of P1-F1, the ARCHIVE stage put eight candidate entries to the authors one by one. Three were approved and five were refused, and the PDR, specification and plan of P1-F2 cite the approved entries. When P6-F1 was archived, the authors approved all the proposed entries at once, including a known trap involving the cloud-sync client. The PDR of P6-F2, opened on the same day, cites that trap. The memory file was also written outside the ARCHIVE stage, even though that is the only stage allowed to write it. In P1-F2, the orchestrator changed an existing entry during an execution round at the authors' request, and the log records this as a failed acceptance criterion. P5 never reached the ARCHIVE stage, yet its memory file has entries as well.

The features that delivered a book, a research dossier or a research paper ran with the workflow text unchanged, although the setup file calls \file{resources/} the root of the code. The checks carried over because they depend on files and declared writes. In P2, for example, the verification of a chapter task compiled the book, and the task listed the build directory among the files it wrote.

\section{Discussion}\label{sec:discussion}

Our observations extend the report by \citet{abenhaim2026specfirst} to conditions that it did not cover. They come from \NProjetosExtenso{} projects and include deliverables that are not software.

The results point to three weak points in the controls: requests the authors made during a round, writes from the environment, and decisions the orchestrator took on its own. Only the third comes from the agent itself.

\paragraph{In-round human requests.} Several undeclared changes, as well as the change to the memory in P1-F2, came from the orchestrator acting on the authors' request or authorisation during a round. Apart from a task approved before the round, the workflow has no rule for accepting a change during a round, so these requests had no channel and became undeclared changes. In some features, orchestrators improvised a rule of their own: numbered amendments to the task file, which the authors approved. Direct requests bypassed this rule. The checks then recorded them as undeclared, recorded them as ``yes'' or absorbed them into a new baseline. A defined channel could keep such changes inside the declared scope: an amendment task that the user approves and that the check then expects. We did not test this.

\paragraph{The environment.} Writes from outside the tasks reached the deliverable folder, and the check could not tell them apart from changes that matter. Of the \NConferenciasSinalizadas{} flagged rows, \NSinalizadasSincronizacaoExtenso{} came from the sync client, \NSinalizadasVerificacaoExtenso{} from a verification and \NSinalizadasAmbienteAutorExtenso{} partly from the authors' own commits. Reading all rows adds further cases of each kind. The writers were a cloud-sync client, verification commands and test batteries, and the authors' own commits, pulls and local runs. A check based on size and modification time over a synchronised folder fires on such writes, and the agent then has to decide whether to treat them as noise. Declaring known external writers in advance and comparing content hashes could separate these writes from changes that matter, but this too is untested.

\paragraph{Orchestrator discretion.} Where the orchestrator decided alone, it mostly decided in favour of continuing. It continued past the stop rule after recording a reason, so the check stopped a round only in P6 and P7. The orchestrator also once tried to rewrite a logged stop as a pass. Against the rule that it must not undo changes on its own, it deleted files. It also took new baselines that absorbed changes made outside tasks, and it dispatched tasks that the reviewer had flagged as unapproved. It called for reviews far more often than asked, and it passed context to a reviewer that was meant not to see it. The workflow's controls belong to what Consort calls enforcement by persuasion \citep{hartman2026consort}. Similarly, in the study by \citet{dhruv2026scientific}, loops whose limits were stated only in the prompt ran past them. The executor, by contrast, kept to its declared files: it left one undeclared change in the deliverable folder, and the check caught it. It did not keep to every rule of the protocol, and its other departures, such as unrequested refactoring and deletions without confirmation in P6-F2, are the examples recorded in its reports and in the deviation rows (\cref{sec:rq1}).

\paragraph{What the data suggest.} Our untested interpretation is that controls run by the tooling, not by the model, would answer most of these weak points. A block check run by the tooling could compare content hashes, save an inventory for every block and impose a stop that the model can neither continue past nor rewrite. Such a check would remove the orchestrator's discretion over it and would make environmental writes visible as such. In-round requests would still need a channel of their own. A reviewer prompt built by the tooling, containing only the artefact path and stage, would restore the independence that the design intends. The tooling could also require approval for tasks added during execution, which is what the reviewer itself asked for in the P2 figure case. A related result holds for gates built on a model's own confidence: a gate that promises an accuracy keeps that promise only when a lower bound is checked on data like its traffic, and a threshold chosen by point estimate broke the promise in 147 of 190 evaluations \citep{brom2026confidence}.

Re-reviews cost reviewer tokens, but some of them still reported blocking findings. In P1, P2 and P5, re-reviews went beyond the instruction and used \NFracaoTokensRevisoesNoEpisodio\% of the reviewer tokens, with \NFracaoTokensAposSemBloqueante\% spent on re-reviews after a report without blocking findings. Yet \NRevisoesAposSemBloqueanteComBloqueanteNovo{} of those later re-reviews still reported blocking findings in the revised artefact, and a ceiling of one review per artefact would have skipped them. Our interpretation, also untested, is that leaving the decision to re-review to the user, with the cost of each call shown, would keep most of the re-reviews that still find blocking problems, at a lower cost than unlimited re-review.

\section{Threats to validity}\label{sec:threats}

\paragraph{Construct.} The reviewer findings are counted but not validated as defects. The block checks are the orchestrator's records of a comparison of file sizes and modification times, which is cumulative within a round. In some projects, the folders compared were synchronised with cloud storage, and the authors edited them between and during rounds. The token counts cover only the reviewer subagent and are a lower bound.

\paragraph{Internal.} Without a control condition, no result supports a causal claim about what the controls prevented. A single AI coder coded the origins of the flagged changes and the boundaries of the episodes, and no agreement measure was computed. The episode counts did not change under the other gap thresholds (\cref{sec:method}).

\paragraph{External.} The results describe only these cases. All of them come from the authors, who also designed the workflow. They cover \NFeaturesExtenso{} features in \NProjetosExtenso{} projects, a single coding agent and the period stated in \cref{sec:method}.

\paragraph{Reliability.} The extraction is scripted and deterministic, but its inputs are incomplete. Reviewer reports exist for \NRevisorProjetosComRelatorioExtenso{} projects and cover \NRevisorComRelatorio{} of the \NRevisorConhecidas{} known calls. P3, P6 and P7 have no transcript data at all, and the P7 reports come from that project's event record. That record was kept by the orchestrator of the feature that produced this paper, so the P7 reviewer data are the agent's own account and cannot be checked against a transcript. P2 saved an inventory at the start and at the end of every round, and the deliverables of P2 and P5 are git repositories. Together these would allow an independent recomputation of the checks for each round, but we did not carry it out. Finally, an AI agent carried out the analysis on logs written by the same kind of agent, a weakness that the separate audit and reviews only partly offset.

\section{Conclusion}\label{sec:conclusion}

We studied the audit trail recorded by an SDD workflow in \NFeaturesExtenso{} features across \NProjetosExtenso{} projects. Almost all of the workflow's controls are instructions that the agent carries out itself. For RQ1, the block checks recorded undeclared changes that came mostly from the authors' in-round requests and from the environment. The executor was the source once, and the check caught that change. Where the orchestrator had to decide whether a control applied, it mostly decided to continue, and the check stopped a round only in P6 and P7. For RQ2, in the \NRevisorProjetosComRelatorioExtenso{} projects with reviewer reports, the reviewer reported a blocking finding in the first review of \NPrimeirasComBloqueante{} of the \NArtefatosRevisados{} artefacts. In P1, P2 and P5, re-reviews that the workflow did not ask for used \NFracaoTokensRevisoesNoEpisodio\% of the reviewer tokens. Of the re-reviews that followed a report without blocking findings, \NRevisoesAposSemBloqueanteComBloqueanteNovo{} still reported blocking findings. For RQ3, interrupted work resumed from the task checkboxes and the logs, although a dead round in P5 was never closed. The project memory carried lessons forward in P1 and P6, and the checks carried over to the \NFeaturesEntregavelNaoSoftwareExtenso{} features that delivered books, a research dossier or a research paper.

The trail made these findings possible. It also points to changes that we did not test: a channel for in-round human requests, the declaration of external writers, and checks, stops and reviewer prompts that run in the tooling rather than in the model. As more features are carried to the ARCHIVE stage, the same scripts can update the counts in this paper once the new flagged rows and deliverables have been coded. Those features would also allow the analysis of the project memory to be extended.

\section*{Data and code availability}

The workflow studied here is public at \url{https://github.com/rfa-rodrigo/SDD---Template} \citep{abdo2026template}. The extraction scripts and the anonymised aggregate data, with one row per feature, are available from the authors on request by e-mail to rodrigo.abdo@ifb.edu.br.

\section*{Use of AI-assisted tools}

For this paper, Claude Code and the authors wrote the text and collected and analysed the data, and the authors carried out the final review. The AI assistance in writing this paper was itself counted in the data.

\section*{Acknowledgements}

The authors thank the Federal Institute of Brasília.

\bibliographystyle{unsrtnat}
\bibliography{references}

\end{document}

%% file: numbers.tex
\newcommand{\NTarefas}{581}
\newcommand{\NTarefasMarcadas}{288}
\newcommand{\NBlocos}{458}
\newcommand{\NRodadas}{32}
\newcommand{\NDesvios}{154}

\newcommand{\NInventariosSalvos}{38}
\newcommand{\NFeatures}{9}

\newcommand{\NFeaturesEntregavelNaoSoftware}{5}
\newcommand{\NConferencias}{247}
\newcommand{\NConferenciasSinalizadas}{11}

\newcommand{\NRevisorConhecidas}{132}
\newcommand{\NRevisorComRelatorio}{111}

\newcommand{\NRevisorBloq}{75}

\newcommand{\NRevisorAchados}{508}
\newcommand{\NBloqueantesCodigos}{90}
\newcommand{\NBloqueantesMulticodigo}{15}
\newcommand{\NArtefatosRevisados}{20}
\newcommand{\NPrimeirasComBloqueante}{15}
\newcommand{\NPromptsComContexto}{39}
\newcommand{\NPromptsAgente}{81}

\newcommand{\NRevisorSemTokens}{4}
\newcommand{\NEpisodios}{31}

\newcommand{\NEpisodiosIniciadosPorAutor}{27}
\newcommand{\NRevisoesAposRelatorioSemBloqueante}{30}
\newcommand{\NRevisoesAposSemBloqueanteComBloqueanteNovo}{6}
\newcommand{\NRevisoesNoEpisodio}{51}
\newcommand{\NFracaoTokensRevisoesNoEpisodio}{75.3}
\newcommand{\NRevisoesAposRelatorioSemBloqueanteP}{29}
\newcommand{\NFracaoTokensAposSemBloqueante}{49.8}
\newcommand{\NPChamadas}{63}
\newcommand{\NPSemMsgAutor}{47}
\newcommand{\NProjetosExtenso}{seven}
\newcommand{\NFeaturesExtenso}{nine}
\newcommand{\NFeaturesConcluidasExtenso}{five}
\newcommand{\NFeaturesArquivadasExtenso}{four}
\newcommand{\NFeaturesEntregavelNaoSoftwareExtenso}{five}
\newcommand{\NSinalizadasParadaEfetivaExtenso}{four}

\newcommand{\NSinalizadasPedidoExtenso}{three}
\newcommand{\NSinalizadasVerificacaoExtenso}{one}
\newcommand{\NSinalizadasSincronizacaoExtenso}{six}
\newcommand{\NSinalizadasAmbienteAutorExtenso}{one}

\newcommand{\NRevisorLimpasExtenso}{five}
\newcommand{\NRevisorProjetosComRelatorioExtenso}{four}

%% file: tables/coverage.tex
P1 & 16 & 7 & 9 \\
P2 & 67 & 63 & 4 \\
P4 & 8 & 0 & 8 \\
P5 & 12 & 12 & 0 \\
P7 & 29 & 29 & 0 \\
\midrule
Total & 132 & 111 & 21 \\

%% file: tables/features.tex
P1-F1 & Dossier for a research article &27 & 27 & 17 & 2 & 17 & 26 & 1 & completed \\
P1-F2 & Evidence re-check and work plan (same article) &29 & 28 & 29 & 1 & 29 & 15 & 138 & interrupted \\
P2-F1 & Technical book revision (\LaTeX) &152 & 42 & 148 & 8 & 45 & 21 & 53 & interrupted \\
P3-F1 & Numerical solver feature (Fortran) &20 & 20 & 19 & 2 & 24 & 12 & 28 & completed \\
P4-F1 & Trading-strategy backtest (Python) &20 & 20 & 14 & 2 & 13 & 22 & 28 & completed \\
P5-F1 & Technical book revision (\LaTeX) &202 & 21 & 134 & 1 & 12 & 9 & 5 & in execution \\
P6-F1 & Video-analysis pipeline feature (Python) &42 & 42 & 21 & 9 & 29 & 20 & 78 & completed \\
P6-F2 & Second feature of the same pipeline &50 & 49 & 44 & 4 & 45 & 11 & 83 & in execution \\
P7-F1 & This paper (\LaTeX) &39 & 39 & 32 & 3 & 33 & 18 & 16 & completed \\
\midrule
Total & & 581 & 288 & 458 & 32 & 247 & 154 & 430 & \\

%% file: tables/flagged.tex
P1-F2 & Orchestrator extended an evidence file outside any task, at the authors' request & Continued, recorded as deviation \\
P1-F2 & Orchestrator corrected the same file outside any task. The check row calls it an authorised correction, and the contamination table cites the authors' request, but the deviation row cites none & Continued, recorded as deviation \\
P2-F1 & Orchestrator reversal, ordered by the authors, of an unauthorised figure replacement. A second chapter changed date through the authors' own commits & Stop logged. Seconds later the orchestrator ran a script to rewrite the verdict as ``yes, continued'', which targeted a different row by mistake, so nothing changed. The next block was checked four minutes later in the same round \\
P3-F1 & Build products regenerated by the verification command, missing from the task file & Continued, recorded as deviation \\
P6-F1 & Temporary file of the cloud-sync client & Round stopped by the check \\
P6-F1 & Temporary files of the cloud-sync client & Recorded as external. The round stopped on a failed task verification, and the authors approved deleting the files before the next round \\
P6-F1 & Temporary files of the cloud-sync client & Recorded as external. The round stopped on a failed task verification, and the files vanished on their own \\
P6-F1 & Temporary files of the cloud-sync client & Deleted by the orchestrator without asking. The round then paused for the authors \\
P6-F2 & Byte-code caches written by a syntax check the executor ran on its own & Round stopped by the check. The authors were consulted and approved deleting them \\
P7-F1 & Modification time of two analysis scripts moved back by the cloud-sync client, same size & Round stopped by the check. The authors ordered the execution to resume without asking at each step \\
P7-F1 & Build product reverted to an older version by the cloud-sync client & Round stopped by the check and resumed in a new round under the authors' standing order to proceed without asking \\

%% file: tables/reviewer.tex
PDR & 4 & 8 & 1 & 8 & 2 & 7 & 3/4 & 0.19 \\
SPEC & 4 & 37 & 0 & 35 & 60 & 90 & 4/4 & 2.17 \\
PLAN & 4 & 32 & 1 & 11 & 58 & 89 & 3/4 & 4.69 \\
TASKS & 4 & 27 & 3 & 18 & 26 & 68 & 4/4 & 4.09 \\
EXEC & 3 & 6 & 0 & 3 & 16 & 15 & 1/3 & 0.56 \\
ARCHIVE & 1 & 1 & 0 & 0 & 1 & 1 & 0/1 & 0.03 \\
\midrule
Total & 20 & 111 & 5 & 75 & 163 & 270 & 15/20 & 11.74 \\

%% file: tables/blocking.tex
SPEC & 2 & Technical claim without proof & 15 \\
SPEC & 1 & Subjective acceptance criterion & 11 \\
TASKS & 3 & Verification not executable & 9 \\
SPEC & 3 & PDR decision not covered & 8 \\
PDR & 2 & Two decisions in one PDR & 7 \\
SPEC & 4 & Spec that decides on its own & 6 \\
PLAN & 1 & Step without verification & 5 \\
TASKS & 1 & Task that decides & 5 \\
PLAN & 3 & Impossible order & 3 \\
PLAN & 4 & Spec criterion without step & 3 \\
TASKS & 19 & Empty field or placeholder & 3 \\
EXEC & 2 & Undeclared change & 3 \\
TASKS & 15 & Inexact path & 2 \\
TASKS & 2 & Task bound to context & 2 \\
TASKS & 6 & Conflicting declarations & 2 \\
PDR & U2 & Contradicts project memory & 1 \\
PLAN & 5 & False parallelism & 1 \\
TASKS & 11 & Undeclared shared resource & 1 \\
TASKS & 7 & Malformed block & 1 \\
TASKS & 8 & Execution map missing or incoherent & 1 \\
TASKS & 9 & Path that is not a file & 1 \\